\documentclass[preprint]{aastex631}
\usepackage{multirow,color}
\usepackage{bbding}
\usepackage{amssymb}
\usepackage{ulem}
\usepackage{verbatim}
\usepackage{url}

\usepackage{graphicx}

\def\gsim{\;\lower4pt\hbox{${\buildrel\displaystyle >\over\sim}$}\;}
\def\lsim{\;\lower4pt\hbox{${\buildrel\displaystyle <\over\sim}$}\;}
\def\grls{\;\lower4pt\hbox{${\buildrel\displaystyle >\over <}$}\;}

\newcommand{\dblind}[1]{{\color[rgb]{0,0,0} will be revealed after the double-blind review process.}}

\shorttitle{GCR Delay due to Late Opening of Solar Magnetic Field}
\shortauthors{Wang et al.}

\graphicspath{{./}{figures/}}

\begin{document}

\title{Cosmic Ray Intensity Variation Lags Sunspot number: Implications of Late Opening of Solar Magnetic Field}

\author{Yuming Wang}
\affiliation{CAS Key Laboratory of Geospace Environment, School of Earth and Space Sciences, University of Science and Technology of China, Hefei 230026, China}
\affiliation{CAS Center for Excellence in Comparative Planetology, University of Science and Technology of China, Hefei 230026, China}
\affiliation{Mengcheng National Geophysical Observatory, University of Science and Technology of China, Mengcheng 233500, China}

\author{Jingnan Guo}
\affiliation{CAS Key Laboratory of Geospace Environment, School of Earth and Space Sciences, University of Science and Technology of China, Hefei 230026, China}
\affiliation{CAS Center for Excellence in Comparative Planetology, University of Science and Technology of China, Hefei 230026, China}

\author{Gang Li}
\affiliation{Department of Space Science and CSPAR, The University of Alabama in Huntsville, Alabama, USA}

\author{Elias Roussos}
\affiliation{Max Planck Institute for Solar System Research, Goettingen, Germany}

\author{Junwei Zhao}
\affiliation{W. W. Hansen Experimental Physics Laboratory, Stanford University, Stanford, USA.}

\correspondingauthor{Yuming Wang \& Jingnan Guo}
\email{ymwang@ustc.edu.cn, jnguo@ustc.edu.cn}

\begin{abstract}
Galactic cosmic rays (GCRs), the highly energetic particles that may raise critical health issues for astronauts in space, are modulated by solar activity with their intensity lagging behind the sunspot number (SSN) variation by about one year. Previously, this lag has been attributed to a combined effect of outward convecting solar wind and inward propagating GCRs. However, the lag's amplitude and its solar-cycle dependence are still not fully understood \citep[e.g.,][]{Ross_Chaplin_2019}. 
By investigating the solar surface magnetic field, we find that the source of heliospheric magnetic field --- the open magnetic flux on the Sun, already lags behind SSN before it convects into heliosphere along with the solar wind. The delay during odd cycles is longer than that during sequential even cycles. Thus, we propose that the GCR lag is primarily due to the greatly late opening of the solar magnetic field with respect to SSN, though solar wind convection and particle transport in the heliosphere also matter. We further investigate the origin of the open flux from different latitudes of the Sun and found that the total open flux is significantly contributed by that from low latitudes where coronal mass ejections frequently occur and also show an odd-even cyclic pattern. Our findings challenge existing theories, and may serve as the physical basis of long-term forecasts radiation dose estimates for manned deep-space exploration missions.
\end{abstract}


\section{Introduction} \label{sec:intro}
Galactic Cosmic Rays (GCRs), which are omnipresent, charged and energetic particles coming from outside of the heliosphere, are affected by the heliospheric magnetic flux as they propagate inward from the heliospheric boundary at about 120 AU  \citep{Krimigis_etal_2013}.
Since decades ago, we have learned that GCR fluxes are constantly affected by variations of the heliospheric magnetic fields, both on short and long time scales. In the short term of days or months, the GCR flux can be altered in the form of Forbush decreases \citep[as first reported by][]{Forbush_1937} due to transient heliospheric structures with more turbulent and intensive magnetic fields such as interplanetary coronal mass ejections \citep[ICMEs,][]{Cane_2000} and stream interaction regions \citep[SIRs,][]{Richardson_2004}.
As GCRs can interact with Earth's atmosphere via ionization processes, such disturbed GCR variations have also been argued to be the link of Sun-climate correlations \citep{Pittock_1978} via changing the global electric circuit and modifying cloud properties \citep{Harrison_etal_2011, Laken_etal_2012, Laken_calogovic_2013}.
In the long term of a few years, the GCR flux was first observed to anti-correlate with sunspot variations \citep{Forbush_1958} since the transport of GCRs is modulated by heliospheric field strength and irregularities that evolve following the quasi-11-year solar cycle \citep{Parker_1965, Potgieter_1998}. 
Specifically, enhanced magnetic flux is more efficient in preventing charged GCR particles from deeply penetrating into the heliosphere, causing decrease of GCR fluxes towards solar maxima. 
The variation of GCR fluxes at Earth has been correlated with various solar and heliospheric parameters, such as the Sunspot Number (SSN), the strength and turbulence level of heliospheric magnetic field (HMF), the heliospheric current sheet (HCS) tilt angle, the open solar magnetic flux, the solar polarity, and so on \citep[][etc.]{usoskin_etal_1998, Cliver_Ling_2001, rouillard_etal_2004, alanko_etal_2007, Potgieter_2013}, and empirical functions describing the GCR dependence on different solar cycle parameters have been proposed \citep[e.g.,][]{Dorman_2001, usoskin_etal_2011, GuoJ_etal_2015}.

In particular, when correlating the GCR and SSN temporal variations, the strongest anti-correlation appears when the GCR profile is shifted backward in time, suggesting a delay of the GCR variation with respect to the solar activity evolution. 
The classic picture to explain this time lag involves the solar wind convection and the GCR transport in the heliosphere \citep{Parker_1965, vanAllen2000, Dorman_2001, Usoskin_etal_2001, Cliver_Ling_2001, Thomas_etal_2014}. That is, GCRs propagate inward throughout the heliosphere and are affected by the magnetic field carried by the outward solar wind during their journey. Meanwhile, different paths of GCRs during different phases of the solar cycle could also influence the arrival times of GCRs.

It has also been observed that the modulation of GCRs has an odd-even cycle dependency \citep{Webber_lockwood_1988, vanAllen2000, Cliver_Ling_2001, Thomas_etal_2014}. This implies that the GCR-SSN lag is much longer (one year or more) during odd solar cycles than that (no more than two months) during even solar cycles.
This has been explained, following the above theory of the cause of the delay, by the different drift patterns and diffusion process of GCR propagation through the heliosphere when the solar magnetic poles have predominantly positive or negative spatial orientation following the 22-year Hale cycle \citep{Jokipii_etal_1995, Jokipii_etal_1977, ferreira_Potgieter_2004}.
Specifically, the polarity of the solar field (often represented by symbol A) is positive/negative when the dominant polar field is outward/inward in the northern hemisphere. The polar field reversal occurs around the time when SSN reaches maximum and divides each solar cycle into A+ and A- half cycles. 
Influenced by the curvature and gradient drift pattern, GCR protons arrive at the inner heliosphere after approaching the solar poles and moving out along the heliospheric current sheet during A+ phase.
In opposition, during A- phase, GCR protons propagate inward along the HCS plane
and leave via the poles. 
Around solar maximum, disturbances in the solar wind are much stronger making it more difficult for particles to propagate inward along the HCS plane.
Odd solar cycles start with A+ polarity and switch to A- so that GCRs experience a slower recovery during the declining phase of the cycle when they predominantly enter the heliosphere along the HCS.
Even cycles (first A- and then A+) experience a faster GCR recovery. 

However, the above GCR transport model alone can hardly predict the GCR-SSN lags at the precision of a few months that could be revealed from observations. 
A recent study \citep{Ross_Chaplin_2019} has shown that during Cycle 24, GCR-SSN lag is about 4 months which is slightly longer than those during preceding even-numbered cycles which were 1-2 months, although not as long as those observed in previous odd-numbered cycles which were longer than a year. 
To explain such exceptions, theoretical models would need to rely on various ad-hoc and adjustable parameters whose values can not be directly verified but are chosen to fit the modeling results with observations. 

In this study, we focus on the solar open field (OF) that forms the large-scale HMF where the GCRs propagate through. Via an investigation of GCR-SSN correlation and the OF-SSN correlation throughout several solar cycles (Cycle 21-24) and at various heliospheric distances (from 1 to about 30 AU), we find that the temporal evolution of GCRs is closely mirroring the temporal variations of the open field which is already delayed with respect to SSN. 
Thus, we propose that the GCR-SSN lag is primarily due to the late opening of the solar magnetic field at the Sun while solar wind convection and GCR transport probably make a secondary contribution to cause a longer delay on top of the OF-SSN delay.
Besides, we find that the OF-SSN delay during odd cycles is longer than that during its sequential even cycles. Instead of the transport effect in the heliosphere, we suggest that the behavior of the open field on the Sun could be the main driver of the odd-even cycle dependence of the GCR-SSN delay.
We further investigate the origin of the OF from different latitudes of the Sun and found that the total OF is largely contributed by that from low latitudes where CMEs are frequently launched from active regions and also show an odd-even cycle dependency as GCR-SSN delay. These findings advance our understanding on the solar cycle evolution and its impact on the heliosphere.

\section{GCR delays at various heliocentric distances }\label{sec:distance}

To clearly show the GCR delay in the heliosphere, we investigate the energetic particle measurements at Earth, Mars, Rosetta swinging around 3 AU, Cassini near Saturn and New Horizon (NH) approaching Pluto during Cycles 23 and 24 (see panels 3 to 7 in Fig.\ref{fig:sc23-24}a for the above 5 GCR datasets). The description of the GCR data is given briefly in the caption of Fig.\ref{fig:sc23-24} and in more details in Appendix \ref{sec:gcr_data}.
We also use the daily SSNs \citep{Clette_etal_2014} which are directly obtained from the SIDC of Royal Observatory of Belgium (\url{http://www.sidc.be/silso/datafiles}; the latest version 2.0 is used), as shown in the first panel of Fig.\ref{fig:sc23-24}a. The second panel shows the OF and will be explained later. 
All of these data are prepared in each Bartels rotation (BR, one BR is 27 days) by calculating the median value of the daily values of these data for each BR. 

\newpage

\begin{figure*}[h]
\begin{center}
\includegraphics[width=\hsize]{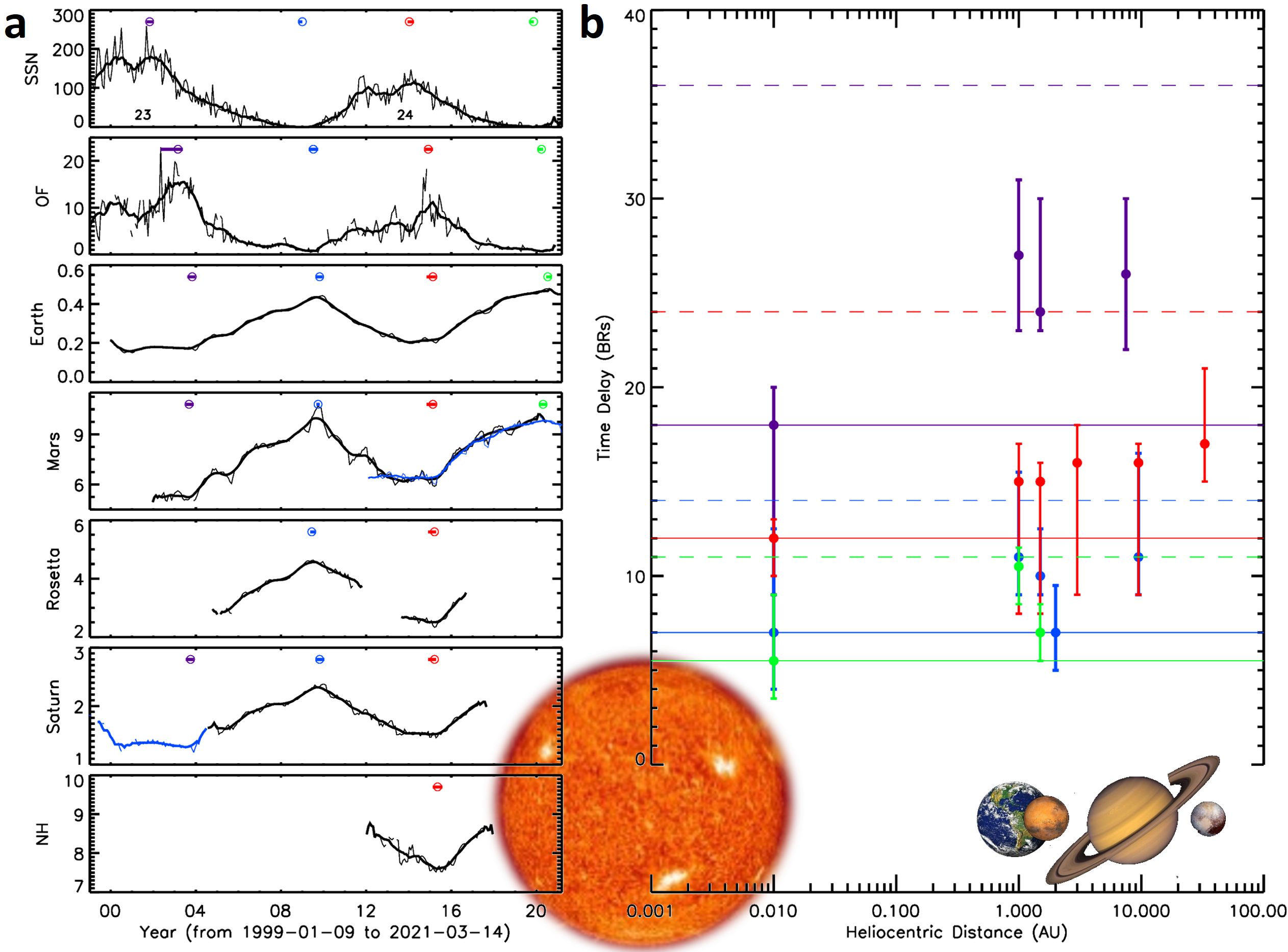}\nonumber
\end{center}
\end{figure*}

\begin{figure*}[h]
\begin{center}
\renewcommand{\baselinestretch}{1}
\caption{\footnotesize{\bf  Time delays of solar open flux and GCR fluxes at five different heliocentric distances with respect to the SSN variation for Solar Cycles 23 and 24.}
{\bf (a)} From the top to bottom, the panels show the solar cycle variations of the SSN \citep[][version 2.0]{Clette_etal_2014}, the open magnetic flux (OF) on the Sun (in units of $\times10^{13}$ Wb, 
derived from the current-sheet source-surface (CSSS) model  \citep{Zhao_Hoeksema_1995} based on the WSO magnetic synoptic charts  \citep{Duvall_etal_1977}, see the main text 
for details), the GCR flux at Earth (measured by the EPHIN \citep{Muller_etal_1995} onboard the Solar and Heliospheric Observatory 
\citep[SOHO,][]{Domingo_etal_1995} of protons above 53 MeV), the GCR flux at Mars ($\sim$ 1.5 AU, measured by HEND  \citep{boynton_etal_2004} onboard Odyssey orbiter of the Martian albedo neutrons with energies 10 eV -- 1 MeV in black and by RAD  \citep{Hassler_etal_2012, Hassler_etal_2014} onboard the Curiosity rover of the Mars's surface dose rates induced by original protons above $\sim$160 MeV in blue  \citep{GuoJ_etal_2018b}), the GCR flux within 4.5 AU 
(measured by SREM \citep{Evans_etal_2008} onboard the Rosetta spacecraft of ions above 49 MeV), the GCR flux near Saturn at about 9.5 AU (measured by Cassini spacecraft of ions above 120 MeV \citep{Roussos_etal_2020}; cruise data are shown in blue until Cassini arrived at Saturn in 2004 July) and the GCR flux from about 22 AU to 40 AU measured by New Horizon (NH, of 120 MeV -- 1.4 GeV ions \citep{Hill_etal_2020}). 
The GCR flux recorded by New Horizon is in units of \#/keV/cm$^2$/sr/sec; the GCR data of RAD are scaled by 32\textmu Gy/day and the GCR fluxes from other instruments are all in the units of the counts per seconds. The temporal resolution of all the data is one Bartels rotation (BR) as shown by the thin curves.
The thick curves are obtained after the running smooth using a 13-BR window. 
The purple, blue, red and green circles
in the first panel mark the four selected extrema (either maximum or minimum) of the SSN, and the circles in the other panels 
mark the positions of the corresponding extrema of the data shown. The horizontal bars across the circles give the uncertainties.   
{\bf (b)} The dots give the time delays for the extrema identified in the left column {\bf a} with the same colors, and the error bars are the uncertainties. The solar distance of 
the open flux is set at 0.01 AU, which is close the cusp surface in the CSSS model where the solar magnetic fields open into heliosphere. The horizontal solid lines mark the open flux delay times of the four extrema, and the dashed lines mark two times of the corresponding delay times. 
}
\label{fig:sc23-24}
\end{center}
\end{figure*}

\newpage

First, we use the peak-alignment method to estimate the GCR time delays with respect to SSN at various distances. 
In this method, we first locate a notable extremum (maximum or minimum) in SSN, and then set a reasonable time range, from $-1$ to $2.5$ yr, to search the corresponding extremum in the parameter of interest, e.g., the GCR flux at Earth. 
The time difference between the two peaks is the time delay. A positive value means the delay and a negative value the lead. 
Since short-term fluctuations due to temporary solar transients may influence the positions of the extrema, we smooth the data over a certain size and repeat the above procedures to obtain the time delay again. 
We change the smoothing window from one BR to 13 BRs, close to one year, and get 13 peak locations for each parameters as well as 13 time delays, of which the median value is adopted as the final time delay (the color circles in Fig.\ref{fig:sc23-24}a or dots in Fig.\ref{fig:sc23-24}b) and the minimum and maximum values are used as the uncertainty (the associated bars in Fig.\ref{fig:sc23-24}).

We select four SSN extrema (two maxima and two minima, the color-coded circles in the top panel of Fig.\ref{fig:sc23-24}a) during Cycles 23 and 24 for the peak-alignment method. It should be noted that during the cycle maxima, SSN depicts a clear double-peak structure (see the solar maxima during Cycle 23 and 24), and the second peak of the SSN around a solar maximum is used in this study to make sure that we have as many extremum values as possible across different datasets which cover various durations (other panels of Fig.\ref{fig:sc23-24}a).

The estimated GCR delays are summarized in Fig.\ref{fig:sc23-24}b, in which the dots with the error bars give the time delays for the extrema identified in Fig.\ref{fig:sc23-24}a with the same colors. It can be found in this panel that the GCR delay is generally longer during solar maxima (marked by purple and red dots) than minima (marked by green and blue dots), 
and could be as long as more than 22 solar BRs or 1.6 years (see the purple dots beyond 1 AU). According to existing particle transport theories \citep[e.g.,][]{Potgieter_2013}, 
such delays are thought to be mainly due to the outward convection of solar cycle conditions of the HMF with the solar wind and the inward diffusive propagation of GCRs. 
Here, we look at this issue from another angle of view. We go to the source of the HMF, i.e., the open magnetic field on the Sun, to see if the delay in respect to SSN already happens before the solar wind outward convection, which may provide an alternative explanation.

Since there is no technique to directly observe solar open field, we use the current-sheet source-surface (CSSS) model \citep[CSSS,][]{Zhao_Hoeksema_1995} to extrapolate and locate the open magnetic fields based on the line-of-sight (LOS) photospheric magnetic fields observed by Wilcox Solar Observatory \cite[WSO,][\url{http://wso.stanford.edu/synopticl.html}]{Duvall_etal_1977}. 
The WSO synoptic charts are stored in Carrington rotations. To obtain the synoptic charts in Bartels rotations, we shift the starting time of the charts and reassemble them. Here, we use WSO data rather than the data from the Michelson Doppler Imager  \citep[MDI,][]{Scherrer_etal_1995} onboard the SOHO spacecraft and the Helioseismic and Magnetic Imager \citep[HMI,][]{Hoeksema_etal_2014} onboard the Solar Dynamics Observatory \citep[SDO,][]{Pesnell_etal_2012} because WSO data cover the past four solar cycles, much longer than the MDI or HMI data. The difference of the extrapolation results between them and the influence on our conclusion could be found in Appendix~\ref{sec_MDIHMI}.

The CSSS model is a widely-used coronal magnetic field extrapolation model \citep[e.g.,][]{Wang_Zhang_2007, Gui_etal_2011}, and compared to the potential field source surface (PFSS) model, it includes the effects of the large-scale horizontal current sheet in the inner corona, the
warped heliospheric current sheet in the upper corona and the volume currents in the outer corona,
and can predict the strength and polarity of HMF better \citep{Zhao_Hoeksema_1995}.
Using this model, we calculate the spherical harmonic coefficients up to 9 orders, and trace magnetic field lines from the solar surface. 
Field lines returning back to the surface are closed field lines, while those crossing over the cusp surface at 2.5 solar radii are open field lines.
Since spatial resolution of the input WSO chart is 72 (in longitude) by 30 (in latitude), we trace one field line from each grid point resulting in a total of $72\times30=2160$ field lines. 
Meanwhile, we calculate the surface area of each grid point. The total area of all the grid points is the total solar surface area. With this information, we can easily derive the open flux, open area and the averaged open magnetic field strength. 
It should be noted that there are sometimes missing data in WSO LOS synoptic charts. If the number of missing data points exceeds 60, i.e., about 3\% of total number of data points, in a BR chart, we simply omit the open field data corresponding to this BR chart. 

We find that the total flux of the open fields shows clear solar cycle variations 
(the second panel of Fig.\ref{fig:sc23-24}a -- OF ) 
and looks well anti-correlated with the GCR fluxes. Using the previous peak-alignment method, we determine the time delay of the open flux after the SSN to be about 18, 8.5, 12 and 5.5 BRs for the four extrema, respectively, as summarized in Fig.\ref{fig:sc23-24}b (see the dots with error bars located at 0.01 AU; the four horizontal solid lines crossing the dots are plotted for the comparison with the time delay of GCRs in the heliosphere). 
The time delay of each GCR dataset is longer than the delay of the corresponding open flux, except for the distance-varying Rosetta data point in blue. 
Considering that the GCR delay may consists of two sources: one is the process in the heliosphere 
including the heliospheric magnetic field convection by solar wind and the particle transport, 
and the other is at the very beginning on the Sun, i.e., the delay of the solar open flux, 
we can find in the figure that the latter may play an important role as none of the estimated GCR delays is longer than two times of the corresponding open flux delay within the orbit of Pluto (indicated by the dashed lines in Fig.\ref{fig:sc23-24}b which correspond to twice the time of the open flux delay marked by the solid lines). 

\section{GCR delays at Earth over the past 45 years}
We further extend the analysis to the past 45 years since May 1976 to March 2021(Cycles 21-24) as shown in Fig.\ref{fig:sc21-24}a, during which
the WSO observations of the solar photospheric magnetic fields are available but continuous GCR observations are only available at Earth 
by ground-based neutron monitors. We use the Oulu data (\url{http://cosmicrays.oulu.fi}) for this study. 
Previous studies show that the delay time can be energy-dependent with shorter delays for higher energy GCRs \citep{ShenZ_etal_2020}, an effect which may be attributed to the energy-dependent transport of GCRs in the heliosphere \citep{Moloto_Engelbrecht_2020}. 
Compared to the SOHO/EPHIN data in Fig.\ref{fig:sc23-24}a, the GCR delay time at Earth is about one to two BRs shorter if the Oulu is used (more details in Appendix \ref{sec:gcr_data} and Fig.\ref{fig:oulu-soho}).

\begin{figure*}[b]
\begin{center}
\includegraphics[width=\hsize]{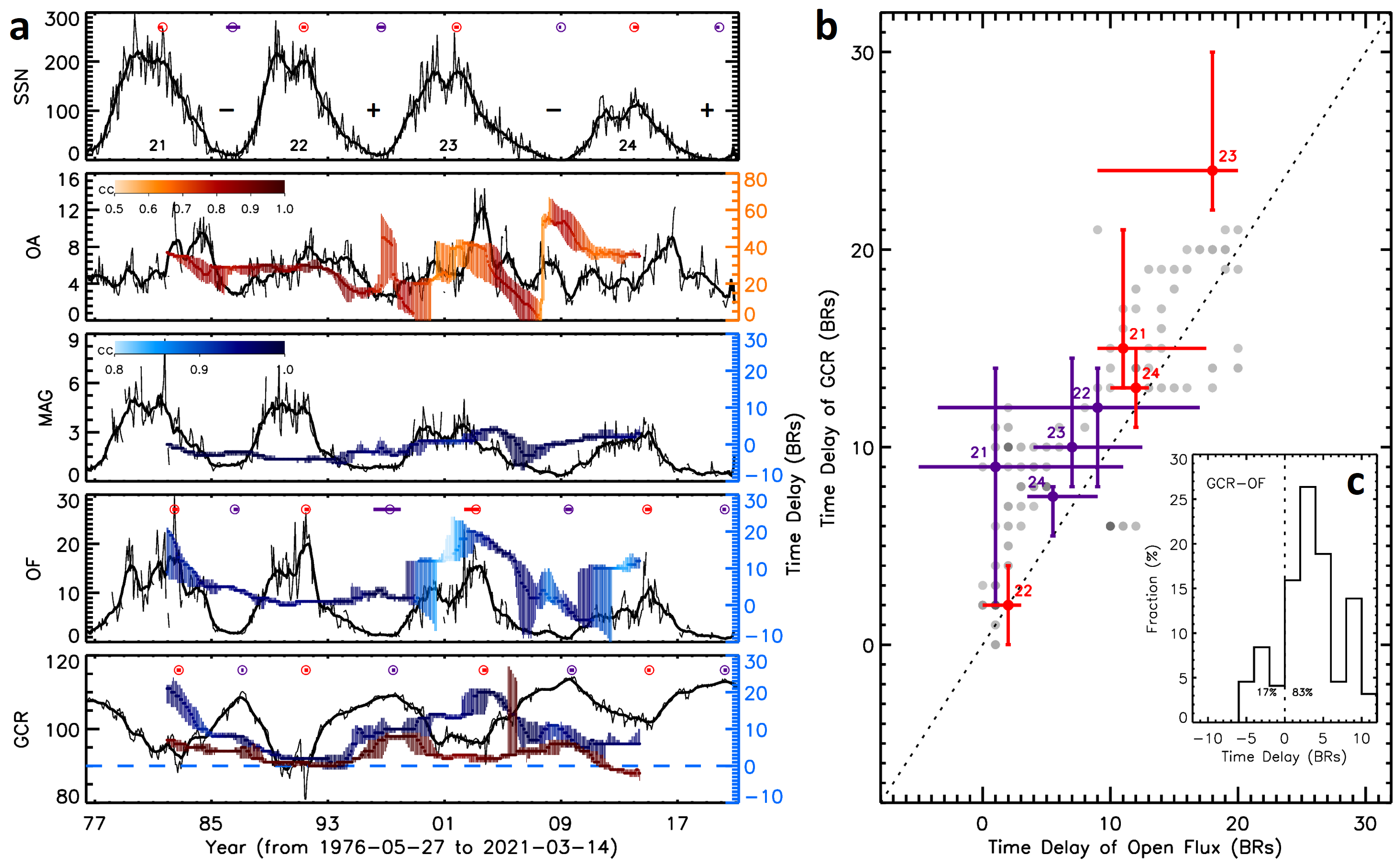}\nonumber
\end{center}
\end{figure*}
\newpage
\begin{figure*}[h]
\begin{center}
	\renewcommand{\baselinestretch}{1}
	\caption{\footnotesize{\bf Time delays of the open magnetic field area, open field strength, open flux on the Sun and GCR fluxes at Earth with respect to the SSN evolution
	for the previous four solar cycles.}
{\bf (a)} From the top to bottom, the panels show the solar cycle variations of
the SSN, open area (OA, in units of \% of total solar surface, TSS), open 
field strength (MAG, in units of Gauss) and open flux (OF) on the Sun and the Oulu monitor count rate at Earth (GCR, counts/sec), respectively, from May 1976 to March 2021. Same as Fig.\ref{fig:sc23-24}a, 
the circles and crossing bars are the selected extrema and their uncertainties. 
The red ones are for the second major maximum of each cycle and the purple ones for 
the solar minima. The color-coded ribbon-like curves in the second through the fifth panels show the solar cycle variations (from 1981 Dec. 12 to 2014 May 26)
of the time delays of the shown parameter relative to the SSN by using cross-correlation analysis on the 13-BR smoothed data.  The last panel has two ribbons with the blue one showing the GCR-SSN delay and the red one for the GCR-OF delay. These ribbons are composed
of dots (the median values of the time delays) and error bars (the 20th/80th percentiles of the time delays),
scaled by the right y-axis of each panel. The colors of the ribbons represent the correlation coefficients, 
scaled by the color bars in the second and third panels. 
{\bf (b)} The red and purple dots with error bars mark the time delays of the open flux and GCRs for the 8 extrema.
The solar cycle numbers are also indicated. 
The gray dots show all the time delays obtained from the cross-correlation analysis. Multiple data points with the same
time delays result in a darker dot. 
The dotted line is the diagonal line.
{\bf (c)} The histogram shows the distribution of the GCR-OF delays
based on the cross-correlation analysis, i.e., the distribution of the red ribbon dots in the last panel in the left column {\bf a}. 
For about 83\% of the data points, the GCR delay is longer than the open flux delay.  
}\label{fig:sc21-24}
\end{center}
\end{figure*}

Using the same peak-alignment method, we select the second major maximum (to keep the consistency with what we did for Cycles 23 and 24) of each cycle and every solar minimum, resulting in a total of 8 extrema (the top panel of Fig.\ref{fig:sc21-24}a), to find the delay of each corresponding extremum in the solar open flux and the GCR flux (the last two panels of Fig.\ref{fig:sc21-24}a).
The results displayed by color dots with error bars and numbers in Fig.\ref{fig:sc21-24}b show that for all the extrema, the GCR delays were roughly equal to (when the dots lie close to the diagonal line) or slightly longer (when the dots lie above the diagonal line) than the corresponding open flux delays and were much shorter than two times of the open flux delays, suggesting that the primary role of the open flux in causing the GCR delay holds for the previous four solar cycles. 
We note that the purple point numbered as 21 (the solar minimum between Cycles 21 and 22) is an exception with the GCR delay much longer. However, the large uncertainties in both time delays make it difficult to discuss this outlier. 

Benefiting from the continuous collection of GCR data at Earth throughout this period, we employ the cross-correlation method to reveal the cycle variation of the delay time.
For any given time, we first select a $N$-year wide segment centering at this time in SSN, and then set the same wide time window on the parameter of interest, e.g., the open flux or the GCR flux, to calculate the correlation coefficient (cc) between the selected segments of the parameter and the SSN. 
The $N$-year window rolls through the parameter over a reasonable time range in order to locate the best correlation within the range. The time delay is the shifted time of the window away from its original position. 
Meanwhile the associated cc value is also obtained. Since the correlation depends on the size of the time window, we change the value of $N$ from 5 to 11 by a step of 0.074 (i.e., one BR), gathering the data of half of a solar cycle to a complete solar cycle, and repeat the above steps to achieve a total of 82 time delays and cc values, among which the time delays associated with a cc less than 0.5 are discarded. 
The median values of these delays and these cc values are chosen as the final time delay and the final cc value, respectively. 
The 20th/80th percentile is chosen as the uncertainty range of the time delay. Applying the above technique to the data throughout Solar Cycles 21-24, we then get the cycle variations of the time delays as shown by the color-coded ribbons in Fig.\ref{fig:sc21-24}a.

\begin{figure*}[b]
\begin{center}
\includegraphics[width=\hsize]{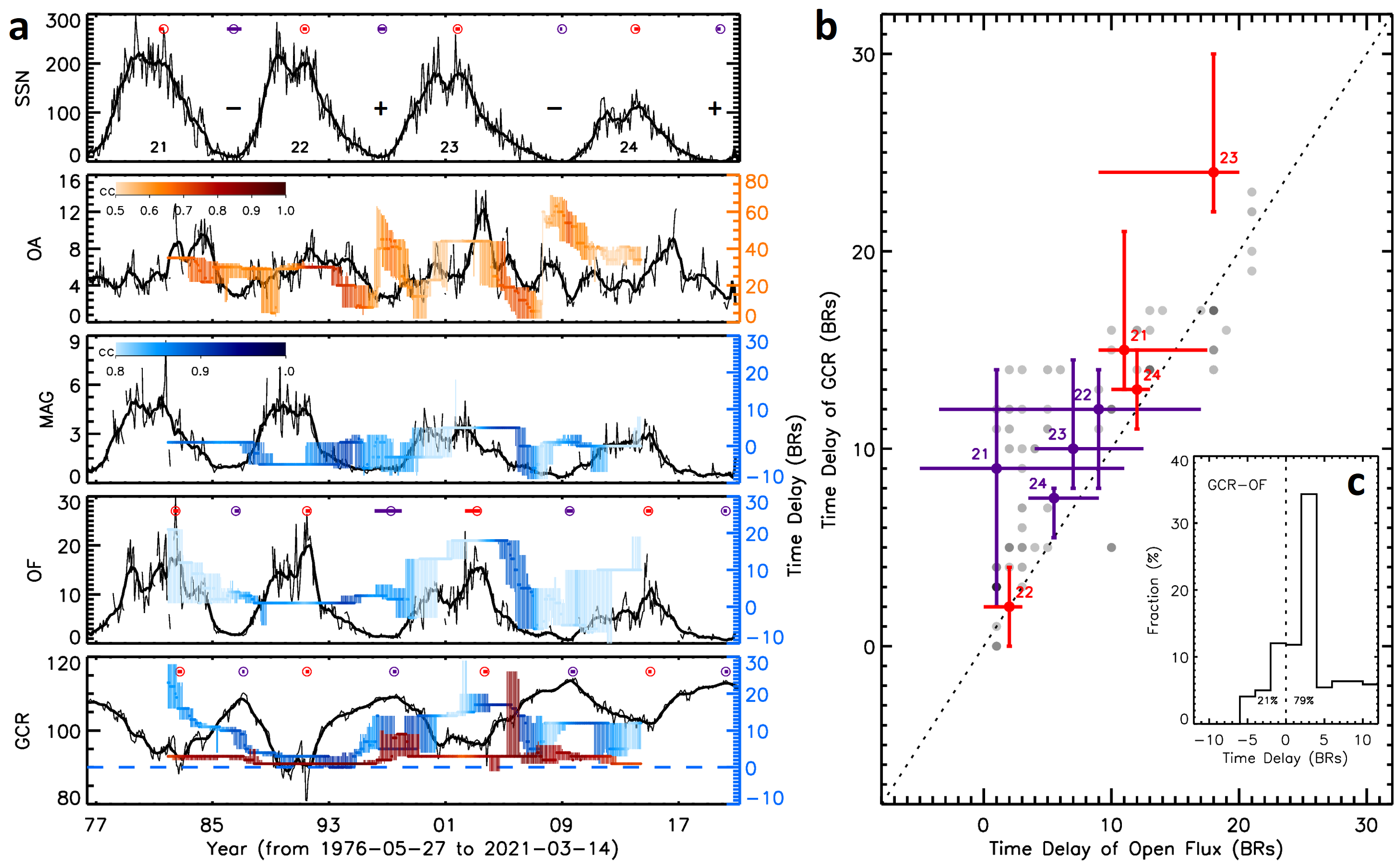}
	\caption{The same as Fig.\ref{fig:sc21-24} except the data used for cross-correlation analysis are not smoothed. 
	}\label{fig:sc21-24_nsm}
\end{center}
\end{figure*}
The smoothness of the data has little influence on the result of the cross-correlation analysis. The color-coded ribbons in Fig.\ref{fig:sc21-24}a are obtained based on the 13-BR smoothed data, and those in Fig.\ref{fig:sc21-24_nsm}a are based on the unsmoothed data. They look similar at the 
scale of a solar cycle. The percentage of the data points with the GCR delay longer than or equal to the open flux delay is 79\%  
(Fig.\ref{fig:sc21-24_nsm}c), 
similar to that based on the smoothed data, which is 83\% (Fig.\ref{fig:sc21-24}c). 
But we also notice that the smoothness of the data is probably an error source for a few gray dots in Fig.\ref{fig:sc21-24}b 
and Fig.\ref{fig:sc21-24_nsm}b, of which the GCR delays are shorter than the open flux delay. 
For example, comparing the red color-coded ribbons in the last panels of Fig.\ref{fig:sc21-24}a and Fig.\ref{fig:sc21-24_nsm}a, we find that during 2013 -- 2014, the GCR flux seems to evolve ahead of the open flux in Fig.\ref{fig:sc21-24}a, but behind of the open flux in Fig.\ref{fig:sc21-24_nsm}a. This illustrates the uncertainty caused by the smoothness of the data sets.

We obtain the cycle evolution of the OF and GCR delays with respect to SSN as shown by the color-coded ribbons, which are scaled by the y-axes on the right, in the last two panels of Fig.\ref{fig:sc21-24}a. 
Especially, the red color-coded ribbon in the last panel (GCR-OF delay) lying below the blue color-coded ribbon (GCR-SSN delay) clearly shows that the GCR delays with respect to SSN are longer than the GCR delays with respect to the open flux. 
The gray dots in Fig.\ref{fig:sc21-24}b represent the GCR-SSN delays (the blue ribbon in the last panel of Fig.\ref{fig:sc21-24}a) versus the OF-SSN delays (the blue ribbon in the second last panel of Fig.\ref{fig:sc21-24}a). Similar to the eight colored dots obtained using the peak-alignment method, the delay time based on the cross-correlation analysis is generally longer for GCR-SSN than for OF-SSN. 
The histogram in Fig.\ref{fig:sc21-24}c further shows the distribution of GCR-OF delay, i.e., dots forming the red ribbon in the last panel of Fig.\ref{fig:sc21-24}a. It shows that about 83\% of the GCR delays are longer than or equal to the open flux delays, and if we only choose the delays with the correlation coefficient $\geq 0.9$, the fraction increases to 93\% (not shown in the figure), supporting again that open flux delay is the main driver for the corresponding GCR lags. 

Meanwhile, we find that the open flux delays during the odd cycles were notably longer than the delays during the sequential
even cycles (see the blue color-coded ribbon in the second last panel of Fig.\ref{fig:sc21-24}a or Fig.\ref{fig:sc21-24_nsm}a). The similar odd-even 
pattern was observed in the GCR delays with respect to SSN \citep{vanAllen2000, Cliver_Ling_2001, Thomas_etal_2014}, but does not appear in the GCR delays with respect to the open flux (see the ribbons in the last panel of Fig.\ref{fig:sc21-24}a or Fig.\ref{fig:sc21-24_nsm}a). 
Thus, the open flux delays could also be a driver of the observed odd-even cycle behavior in the GCR lags, that has been solely attributed to the different drift patterns of GCR propagation through the heliosphere following the 22-year Hale cycles, i.e., GCRs drift in through the poles during A+ cycle or along the HCS close to the ecliptic plane during A- cycle as previously reported \citep{Jokipii_etal_1995, ferreira_Potgieter_2004, Potgieter_1998, Potgieter_2013}. 
According to the analysis here, the drift effect due to the solar polarity changes is perhaps only a secondary effect in driving the global variation of GCRs.

\section{Origin of the open flux delay on the Sun}
\begin{figure*}
\begin{center}
\includegraphics[width=\hsize]{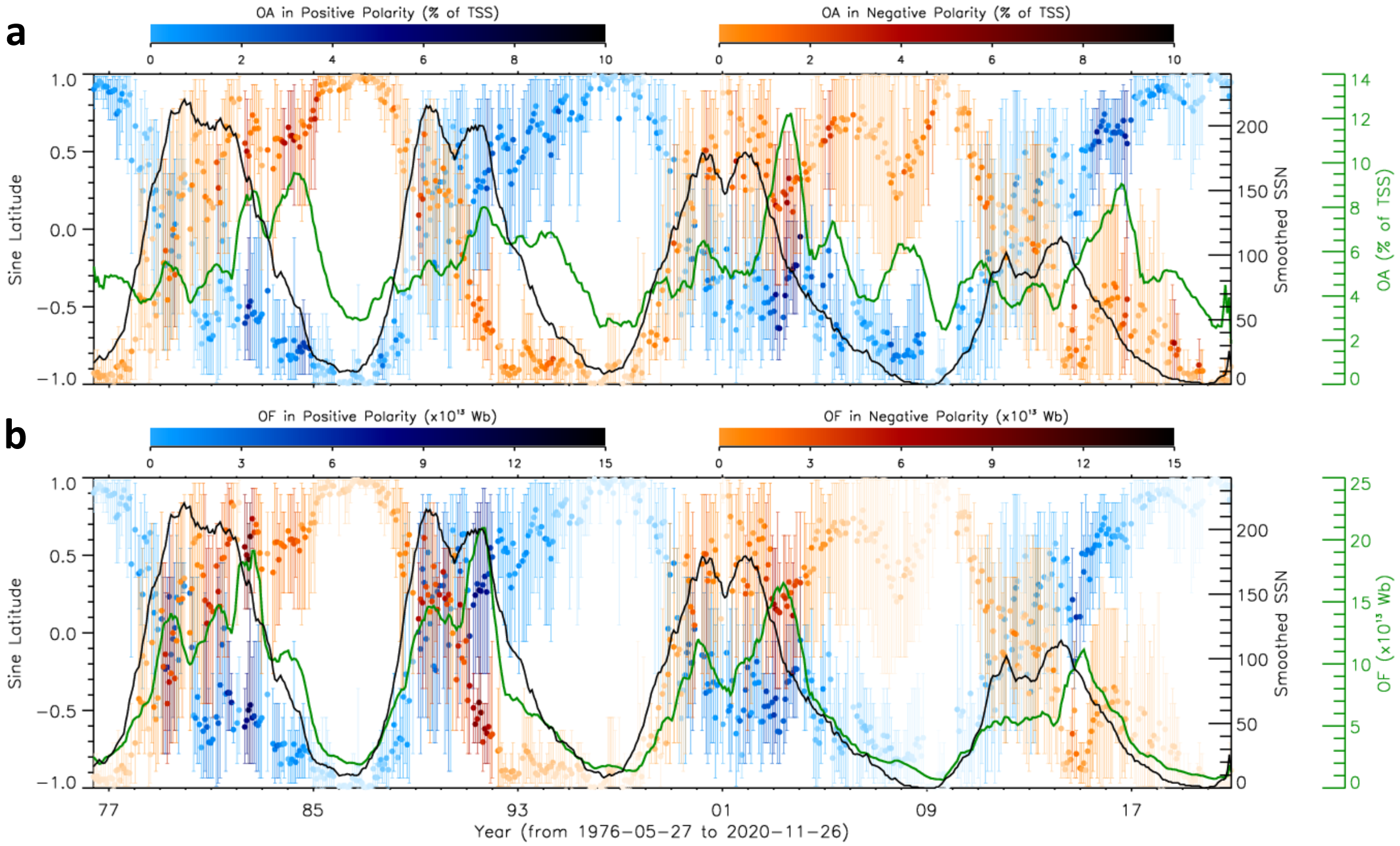}
\caption{{\bf Cyclic diagrams of the solar open area and open flux.} 
{\bf (a)} The latitudinal and size evolution of the open area with the solar cycles. The blue/red dots indicate the latitudinal 
centroid of the positive/negative open field lines. The darker the color, the bigger the total area, which is scaled by the color bars on top of the panel. The error bars give the 20th/80th percentile of the open field lines in latitude.   
The smoothed SSN (black curve) and total open area (green curve) are overplotted.
{\bf (b)} Similar to {\bf (a)}, but for the open flux.
	}\label{fig:open_field}
\end{center}
\end{figure*}

To further find out the origin of the delay of the open flux variation with respect to SSN variation, we investigate the area (OA) and the averaged magnetic field strength (MAG) of the solar open fields (the second and third panels of Fig.\ref{fig:sc21-24}a or Fig.\ref{fig:sc21-24_nsm}a), whose product is the open magnetic flux (OF, shown in the fourth panel). 
Using the same cross-correlation analysis, we find that the MAG changes with the SSN almost synchronously with a small delay varying between $\pm3$ BRs (the blue-coded curve), but the OA lags behind the SSN substantially by about 31 BRs or 2.3 years on average, sometimes by more than 40 BRs corresponding to 3 years (see the red-coded curve). 
Such great delays of the open area cause the open flux to significantly lag behind the SSN. Note that the delay of open area around the solar minimum between Cycles 23 and 24 changed abruptly. 
This segment is not reliable, because during that period there were multiple local maxima and minima even in the smoothed open area data, making it difficult to determine a unique association between the SSN minimum and the open area minima.

The detailed evolution of the open area and flux over solar cycles is shown in Fig.\ref{fig:open_field}a and b, respectively. The open field of positive or negative polarities appears near one polar region at solar minima, gradually migrates toward low latitudes with increasing solar activity 
until solar maxima, and then continuously moves to the other polar region. During the course, 
the latitudinal variations of the total open area and the open flux follow the SSN, but their amplitudes evolve with a lag behind the SSN and reach the maximum generally during the declining phase of each cycle.

\newpage

\begin{figure*}
\begin{center}
\includegraphics[width=\hsize]{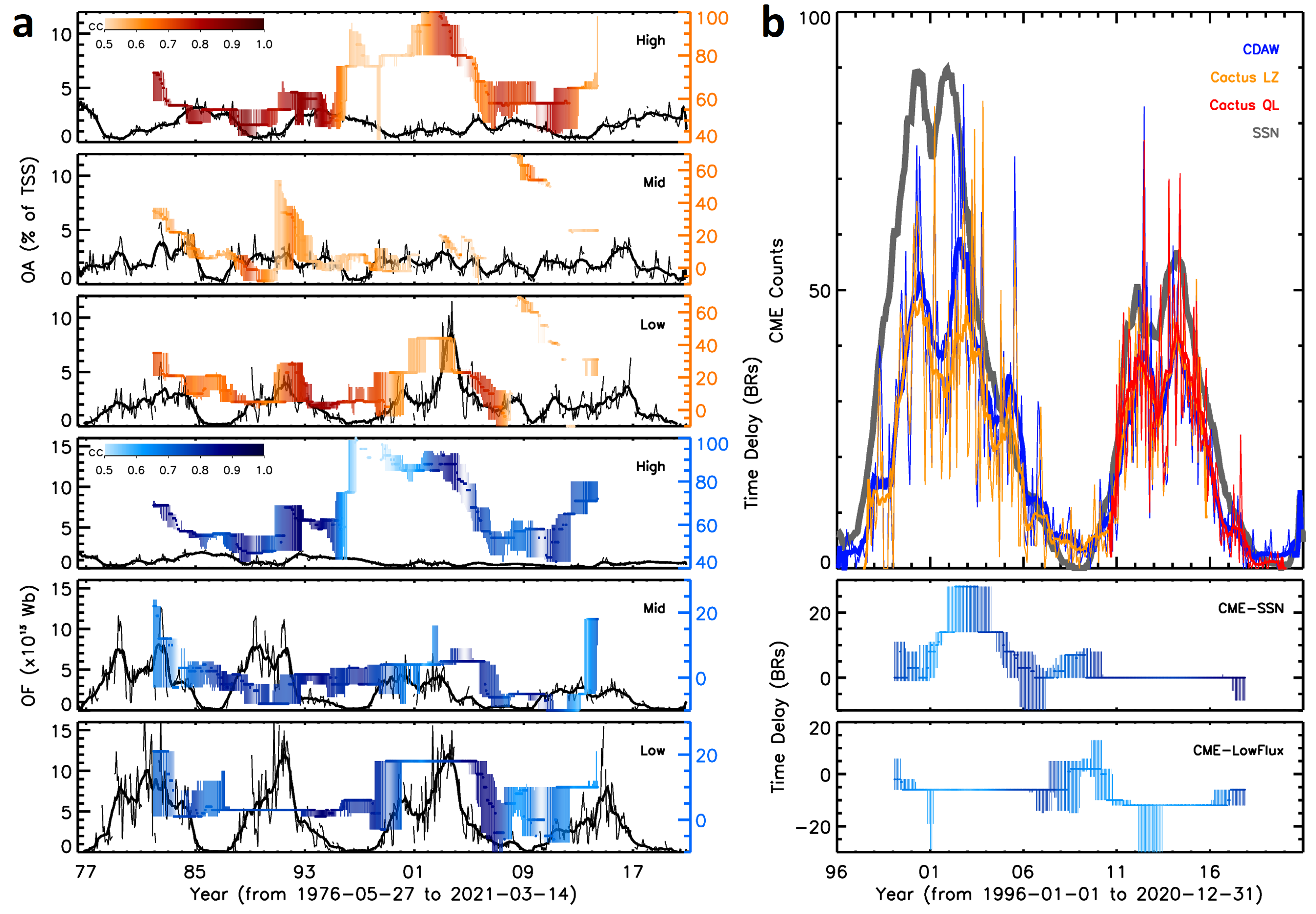}
\caption{{\bf Solar cycle variations of the high, mid and low latitudinal open fields and the CME occurrence.}
{\bf (a)} From top to bottom are the open areas at the high (beyond $\pm60^\circ$), mid (between $\pm30^\circ$ and $\pm60^\circ$) and low (within $\pm30^\circ$) latitudes and the open fluxes at the the high, mid and low latitudes. 
The color-coded ribbons have the same meaning as those in Fig.\ref{fig:sc21-24}a, showing the delays relative to the SSN.
{\bf (b)} The top panel shows the number of CMEs per BR with a speed larger than $500$ km s$^{-1}$. Thin curves are the original counts and thick ones are the 13-BR smoothed values. CME counts derived from CDAW CME catalog \citep {Yashiro_etal_2004} are indicated in blue and those from Cactus catalog \citep {Robbrecht_Berghmans_2004} are in orange and red. The smoothed SSN is overplotted as the gray curve (value is scaled by $0.4$).
The bottom two panels show the time delays of the CDAW CME counts relative to the SSN and low-latitudinal open flux, respectively. The color bar in the fourth panel of {\bf (a)} is used.
	}\label{fig:2comp}
\end{center}
\end{figure*}

Solar open field originates mainly from coronal holes \cite[normally located at high latitudes,][]{WangY_etal_1996} and sometimes from active regions \cite[distributed at low latitudes,][]{Schrijver_Derosa_2003}. 
We then investigate the open area (OA) and open flux (OF) rooted in the high-(beyond $\pm60^\circ$), mid-(between $\pm30^\circ$ and $\pm60^\circ$) and low-(within $\pm30^\circ$) latitudinal zones, as displayed in Fig.\ref{fig:2comp}a. 
It is found that the open areas in different latitudes are on the same order of magnitude. The variations of the open areas at low and mid-latitudes follow the SSN with a delay, and that in polar regions are more or less reversed in phase.
Consequently, the phases of the open flux at high and low latitudes are reversed too. This can be understood by noticing that the solar open field is mainly contributed by the lowest-order of the multi-polar magnetic fields \citep{WangY_etal_2005, JiangJ_etal_2011}, i.e.,
the axial dipole field and the equatorial dipole field that provide most of the open fluxes at high and low latitudes, respectively, with reversed phases. 

The open flux at low latitudes should be paid attention as it is overall about two times of the high-latitude open flux.  
Considering that low latitudes are the main locations of active regions that can also contribute to the heliospheric magnetic flux through coronal mass ejections \citep[CMEs,][]{Luhmann_etal_1998, Owens_Crooker_2006}, we check the solar cycle variation of the CME occurrence number and its relationship with the low-latitude open flux. 

\begin{figure*}
\begin{center}
\includegraphics[width=0.8\hsize]{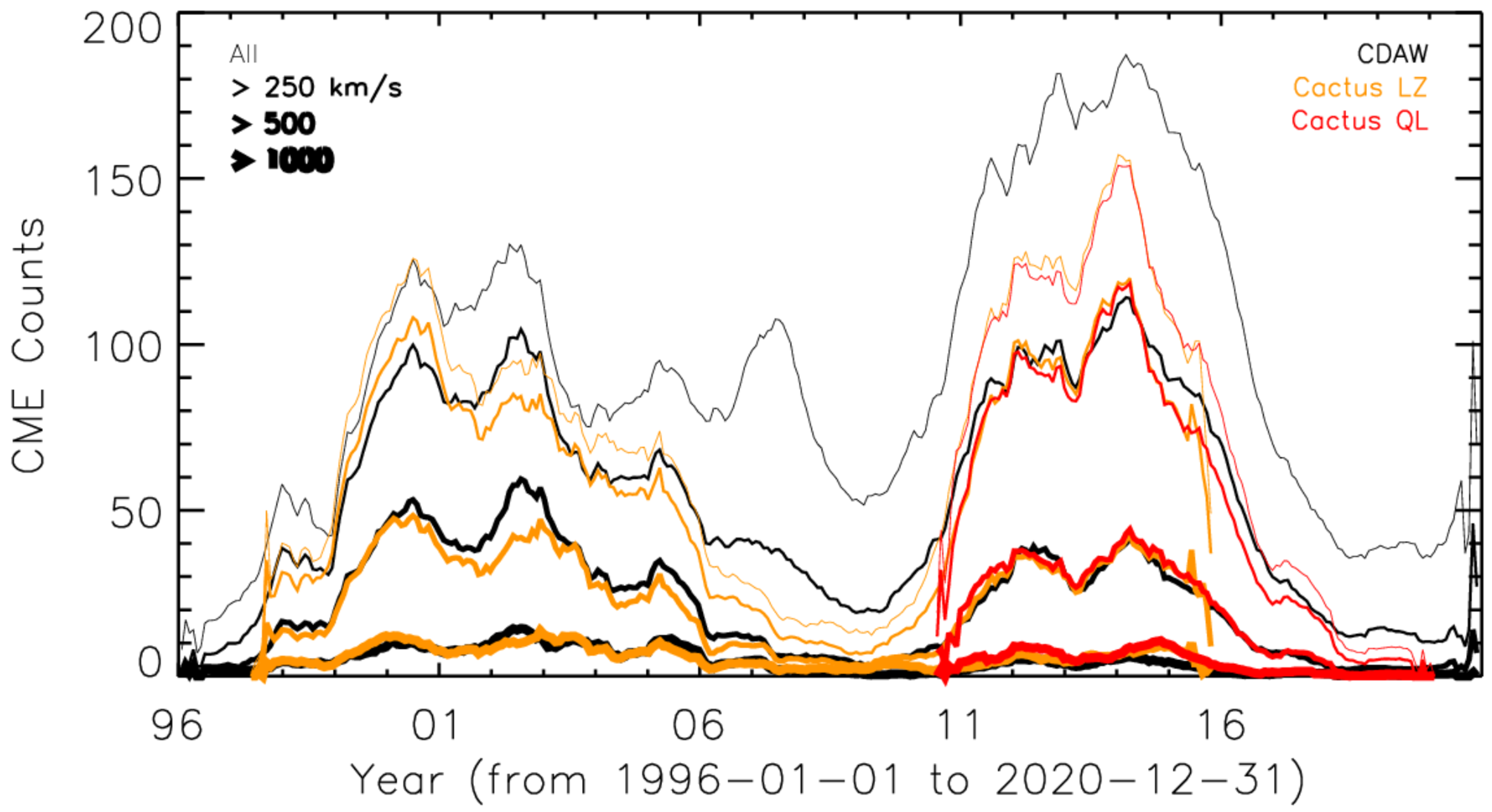}
	\caption{The solar cycle variations of the CME numbers according to the CDAW CME catalog \citep{Yashiro_etal_2004} (black) and the Cactus catalog  \citep{Robbrecht_Berghmans_2004} based on the LASCO level zero data (orange) and on the LASCO quick look data (red). From thin to thick, the lines show the numbers of CMEs with the speed larger than 0, 250, 500 and 1000 km s$^{-1}$, respectively.
	}\label{fig:CMEs}
\end{center}
\end{figure*}

 CMEs are continuously observed by the Large Angle and Spectrometric Coronagraph \citep[LASCO,][]{Brueckner_etal_1995} on board the SOHO spacecraft since 1996. There are several CME catalogs covering Cycles 23 and 24. Here we choose one manually-maintained CME catalog and one machine-automated catalog. The CDAW catalog (\url{https://cdaw.gsfc.nasa.gov/CME_list/}) is manually-maintained \citep{Yashiro_etal_2004}, and the Cactus catalog (\url{http://sidc.be/cactus/}) is based on a machine-automated algorithm \citep{Robbrecht_Berghmans_2004} using either the level zero (LZ) data (until 2015-10-31) or the quick look (QL) data (from 2010-07-09). 
The CME linear speed in the LASCO field of view is provided in the CDAW catalog and the median velocity is provided in the Cactus catalog. They are mostly comparable for CMEs without significant acceleration. 
CME counts with a certain speed threshold is calculated for each BR as shown in Fig.\ref{fig:CMEs}. The comparison of the two different catalogs shows a better consistency with a larger threshold. 
As a result, we choose the CMEs with a speed larger than 500 km s$^{-1}$ for analysis (Fig. \ref{fig:2comp}b) as the two catalogs are consistent and the CME number is large enough. 

It is remarkable that the CME occurrence rate also shows the odd-even solar cycle pattern as revealed by the cross-correlation analysis (the top and middle panels of Fig.\ref{fig:2comp}b), i.e., it notably lagged behind the SSN in Cycle 23, but changed synchronously with the SSN in Cycle 24.
Previous studies presented a model for dynamical energy balance in the flaring solar corona which predicted a time lag between flare occurrence and the supply of energy to the corona \citep{Wheatland_Litvinenko_2001}. 
Studies of solar flare rates and SSN over 5 solar cycles confirmed this time delay and also found it much longer for odd solar cycles than for even cycles \citep{Temmer_etal_2003}. 
Both flares and CMEs represent the process of the gradual accumulation and impulsive release of the magnetic energy in the solar corona. This process may explain the flare/CME lagging behind the SSN, but not for their odd-even behaviors, which deserves further studies.

We further compare the CME numbers with open flux at low latitudes (the last panel of Fig.\ref{fig:2comp}b), and find that the open flux lagged behind the CME rate almost constantly, by about 6 BRs in Cycle 23 and 10 BRs in Cycle 24. This implies that part of the extrapolated solar open flux should result from the reconfigured magnetic fields after CMEs, consistent with the previous studies that up to 30\% heliospheric field could come from active regions \citep{Schrijver_Derosa_2003} and CMEs could contribute to the open solar magnetic fields \citep{Luhmann_etal_1998}. According to the time delay between CME counts and open flux in Fig.\ref{fig:2comp}b, we conjecture that the time scale of the eruption-driven field opening that can be reflected in the photoshperic magnetogram and the extrapolation is about half a year. This finding sheds light on the origin and opening of the solar magnetic flux at low latitudes.

\section{Conclusion}\label{sec:conclusion}

In summary, by investigating the GCR fluxes at various heliocentric distances and the solar open flux over the previous 
four solar cycles, we found that the GCR-SSN delay is longer than the OF-SSN delay, and both of them show the same odd-even cyclic pattern.
We further found that the open flux delay is mainly due to the significant delay of the solar magnetic field opening, and CMEs make important contributions to the open flux at low latitudes.
We also showed that the frequency of CME occurrence has an odd-even cycle dependency. 
Considering that the solar open flux is the source of HMF, we conclude that the GCR-SSN delay has its major origin on the Sun and is dominated by the solar cycle evolution at and below the solar surface. In other words, the delay of the open flux on the Sun is the primary contributor to the GCR delay in the heliosphere, and consequently is also the major cause of the odd-even cycle pattern of the GCR delay. This finding adds a new dimension (open flux delay) to the existing theories that address the GCR-SSN delay mainly based on the solar wind convection and transport of GCRs in the heliosphere, and should be folded into existing theoretical models.
The correlations established among SSN, open flux and GCR intensity in this study provide a basis for the long-term forecast of the GCR radiation levels in the heliosphere which are important concerns for future human space missions \citep{cucinotta2017}.


\begin{acknowledgments}
We acknowledge the use of the Sunspot number data from SIDC of Royal Observatory of Belgium,
the synoptic charts of photospheric magnetic field from Wilcox 
Solar Observatory, the SOHO/MDI and SDO/HMI  instruments, the GCR count rates from the 
Cosmic Ray Station at Oulu, the energetic particle
data from the Odyssey/HEND, the MSL/RAD, the Rosetta/SREM,
the New Horizon/PEPSSI and the Cassini MIMI/LEMMS experiment, and the CME occurrence numbers from the CDAW catalog and the Cactus catalog. We thank Beatriz Sanchez-Cano, Cary Zeitlin, and Matthew E. Hill for the advice on the data usage. We also thank Jie Jiang for valuable discussion about the solar cycle evolution of the magnetic field on the Sun.
This work is supported by the Strategic Priority Program of the Chinese Academy of Sciences (No.XDB41000000) and the NSFC (Nos 42188101 and 42130204). J.G is also supported by NSFC (No.42074222). Y.W. is particularly grateful to the support of the Tencent Foundation.
\end{acknowledgments}



\setcounter{figure}{0}    
\renewcommand{\thefigure}{A\arabic{figure}}
\setcounter{table}{0}    
\renewcommand{\thetable}{A\arabic{table}}
\setcounter{equation}{0}    
\renewcommand{\theequation}{A\arabic{equation}}
\appendix

\section{GCR Data sets}\label{sec:gcr_data}

\begin{figure*}
\begin{center}
\includegraphics[width=0.8\hsize]{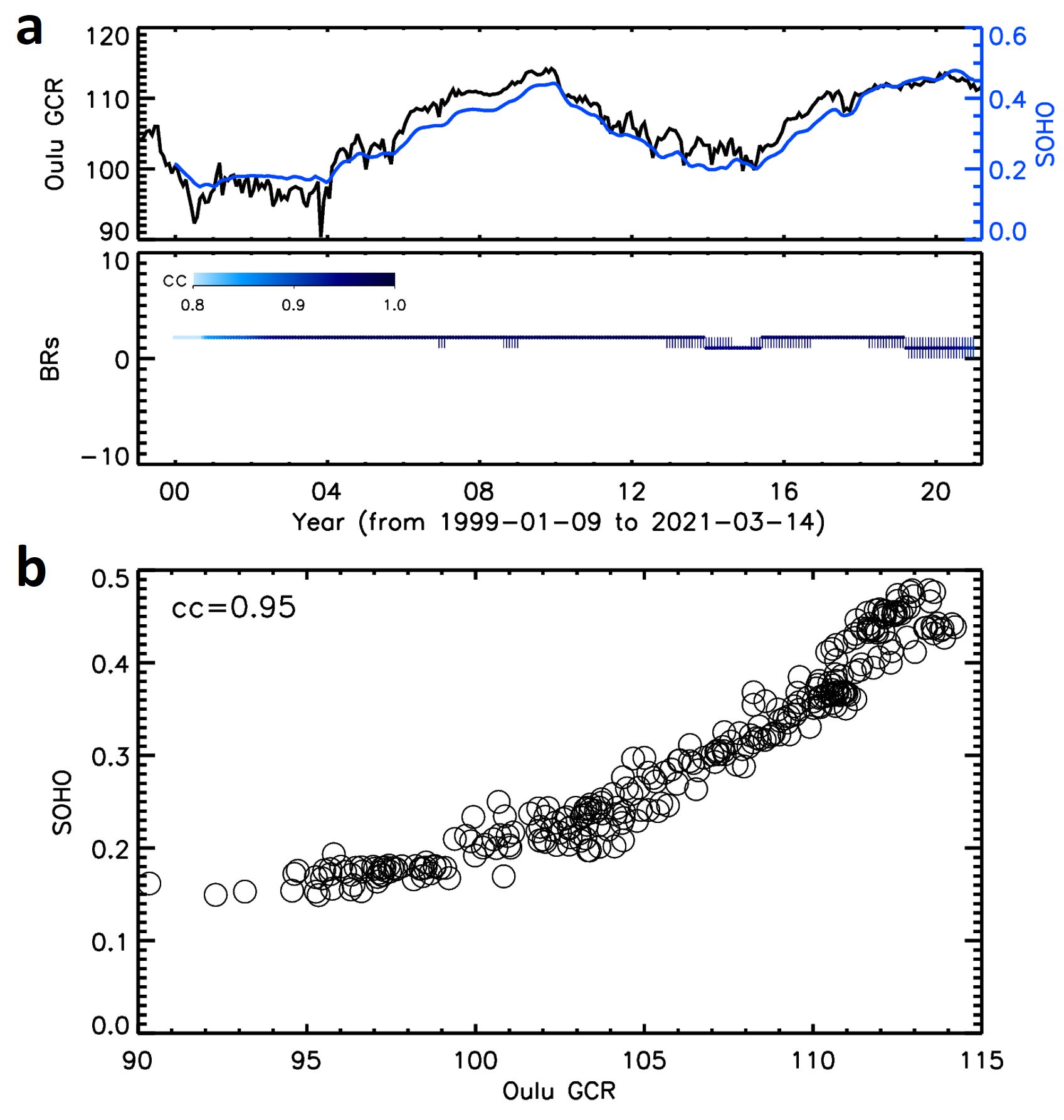}
\caption{The upper panel of (a) shows the comparison between the BR data of the Oulu GCR flux (black) and that of the SOHO/EPHIN energetic particle flux (blue). The lower panel of (a) shows the time delay of the SOHO/EPHIN data with respect to
the Oulu GCR data based on the cross-correlation analysis, with the same meaning as that shown in Fig.\ref{fig:sc21-24}a. Panel (b) shows 
how well the Oulu data and the SOHO/EPHIN data are correlated when 2-BR shift is applied.
	}\label{fig:oulu-soho}
\end{center}
\end{figure*}

The GCR data at Earth in Figure~\ref{fig:sc23-24} are measured by the Electron Proton Helium Instrument (EPHIN) which
is part of the Comprehensive Suprathermal and Energetic Particle Analyzer \cite[COSTEP,][]{Muller_etal_1995} on the SOHO spacecraft. 
As a proxy for the measurement of GCRs, we utilize the channel for particles penetrating through the last detector with minimum ion energies of 53 MeV/nuc. 
The GCR data at Earth in Figure~\ref{fig:sc21-24} and \ref{fig:sc21-24_nsm} are measured by the neutron monitor at Oulu and the averaged daily values are downloaded from \url{http://cosmicrays.oulu.fi}. 
As GCRs enter Earth's atmosphere, they collide with atmospheric particles, producing secondary particles such as neutrons, which are then observed at the detectors situated around the globe. 
Ground level enhancement induced by solar energetic particles (SEPs) are removed and the daily data are binned into each BR with the median values used in the following analysis. 
The Oulu neutron monitor has been recording data since 1964. Given the Earth's geomagnetic field and atmospheric shielding, the cut-off rigidity for Oulu is 1 GV, 
corresponding to about 430 MeV, much higher than that by SOHO/EPHIN. Thus, the delay time of the Oulu GCRs will be shorter than SOHO particles as illustrated
in Fig.\ref{fig:oulu-soho}a, in which the upper panel shows the comparison between the BR data of the Oulu GCR flux
and that of the SOHO/EPHIN energetic particle flux and the lower panel the time delay of the SOHO/EPHIN data with respect to the Oulu GCR data based on the cross-correlation analysis, with the same meaning as the ribbons in Fig.2a. It is found that the amplitude of the time shift between the two data sets is steadily one or two BRs. 
When 2-BR shift is applied, the Oulu data and the SOHO/EPHIN data are well-correlated with the correlation coefficient as high as 0.95 (see Fig.\ref{fig:oulu-soho}b).

The GCR data at Mars include the data from the High Energy Neutron Detector \citep[HEND,][]{boynton_etal_2004} onboard the Mars Odyssey spacecraft (\url{https://pds-geosciences.wustl.edu/missions/odyssey/grs_edr.html}) and the data from the Radiation Assessment detector \citep[RAD,][]{Hassler_etal_2012} onboard the Curiosity Mars rover (\url{https://pds.nasa.gov/ds-view/pds/viewDataset.jsp?dsid=MSL-M-RAD-3-RDR-V1.0}). HEND measures the albedo neutrons that are generated by primary GCRs in the Martian environment and scattered upwards to the orbit. The detector with the thickest moderator layer (about 30 mm) is most sensitive to the neutrons with energies 10 eV--1 MeV and its count rate is used in this study. RAD data are the absorbed dose rate, which is the energy deposit rate by all surface GCRs (including both primary and secondary particles generated in the atmosphere) in the plastic detector. SEPs are removed from both datasets before the daily values are averaged into those for each BR (median values are used).

The data of the Standard Radiation Environment Monitor \citep[SREM,][]{Evans_etal_2008} aboard the Rosetta  \citep{Glassmeier_etal_2007} are available at \url{https://spitfire.estec.esa.int/ODI/dplot_SREM.html}. The TC2 channel with the energy threshold of about 49 MeV for protons is used in this study. SEPs are removed and daily values are binned into BR-averaged values. It should be noted that the distance of the Rosetta spacecraft swang greatly between 1 and 4.5 AU after 2009  \citep{Honig_etal_2019}, and therefore its GCR profile has the spatial gradient effects due to the varying heliospheric distance. 
The same effect applies for the cruise-phase Cassini spacecraft (blue line in the sixth panel of Fig.\ref{fig:sc23-24}a, from 1 AU to 9.5 AU before July 2004) and the New Horizons data (between 22 and 40 AU). The gradient of the GCR flux is only about 2\%--4\% per AU (at distances $<$ 10 AU) according to multi-spacecraft observations  \citep{Honig_etal_2019, Roussos_etal_2020}. The heliospheric distance for each dot shown in Fig.\ref{fig:sc23-24}b is marked using the distance of the spacecraft when the extremum is identified and extra caution should be taken for those obtained from these distance-changing datasets.

The data from the Cassini spacecraft were obtained through the Low Energy Magnetospheric Measurement System (LEMMS), one of the three energetic particle detectors of the Magnetosphere Imaging Instrument (MIMI) suite  \citep[][\url{https://pds-ppi.igpp.ucla.edu/search/view/?f=yes&id=pds://PPI/CO-S-MIMI-4-LEMMS-CALIB-V1.0}]{Krimigis_etal_2004}. LEMMS is a double-sided energetic particle telescope, with 57 counters designed to measure electrons and ions above several 10 keV and up to about 10 MeV (electrons) and several 100 MeV/n (ions). GCR protons penetrate the shielding of LEMMS and get recorded as a low intensity noise signal, that can be isolated from other magnetospheric or solar wind particle sources, as described in detail in earlier studies  \citep{Roussos_etal_2020}. The measurements shown here are from LEMMS channel E6 that its GCR response is dominated by $>$120 MeV protons.

The Pluto Energetic Particle Spectrometer Science Investigation   \citep[PEPSSI,][]{Mcnutt_etal_2008} onboard New Horizon \citep{Stern_2009} were designed to measure keV--MeV pick-up ions from Pluto's outgassing atmosphere. 
Recently, the channels which were originally used to detect $<$MeV Jovian electrons have been successfully used to approximate the deep space GCR fluxes \citep{Hill_etal_2020} and the data are described and accessible here \url{http://sd-www.jhuapl.edu/pepssi/analysis/reducedData/}. The channel corresponding to ions between 120 MeV and 1.4 GeV is used in this analysis. A few SEP events are removed and daily values are binned into BR-averaged values.

\section{Comparison between WSO and HMI/MDI synoptic charts}\label{sec_MDIHMI}
Compared to the magnetograms from space-borne instruments,
e.g., the SOHO/MDI and the SDO/HMI, which provide the synoptic charts with the resolution of 3600 by 1080 and 3600 by 1440 in Cycles 23 and 24, respectively, the spatial resolution of WSO data is too low. However, MDI data just covered the period from 1996 to 2010, and HMI data from 2010 to date. 
Thus, we use WSO data for the long-term analysis here. 
But we also check if the low resolution of WSO data has any influence on the coronal magnetic 
field extrapolation and consequently affects the statistical results. To do so, we apply the same procedures on the MDI and HMI synoptic charts (\url{http://soi.stanford.edu/magnetic/index6.html} and \url{http://jsoc.stanford.edu/HMI/LOS_Synoptic_charts.html}, respectively) to derive the open flux, open area and the open field strength. 

\begin{figure*}[b]
\begin{center}
\includegraphics[width=\hsize]{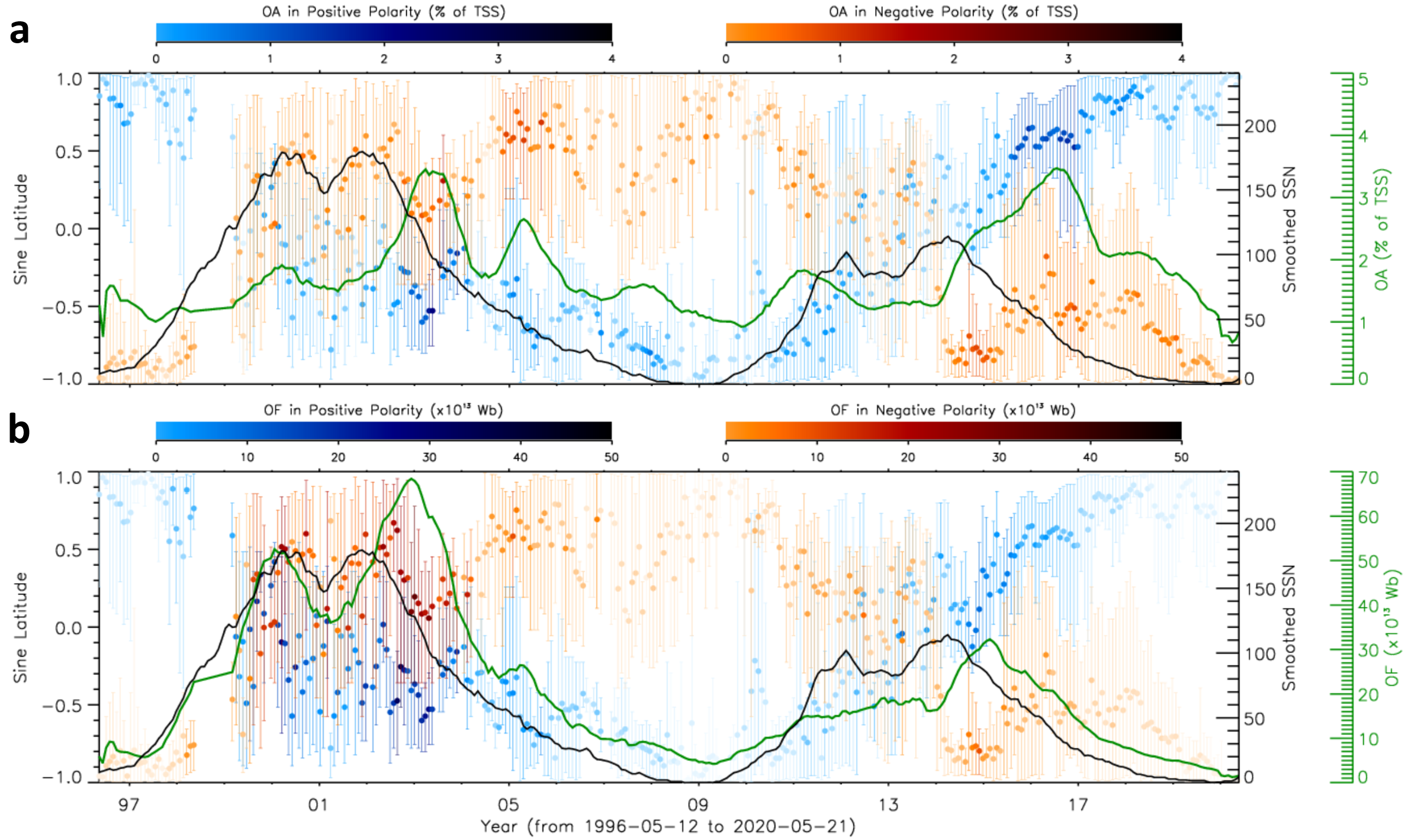}
	\caption{Similar to Fig.\ref{fig:open_field}. The data points before 2011 June 25, which is the beginning of BR 2414, are 
	derived based on the MDI synoptic charts, and those after 2011 June 25 are obtained based on the HMI synoptic charts. 
	}\label{fig:MDI_HMI}
\end{center}
\end{figure*}

\begin{figure*}[b]
\begin{center}
\includegraphics[width=0.5\hsize]{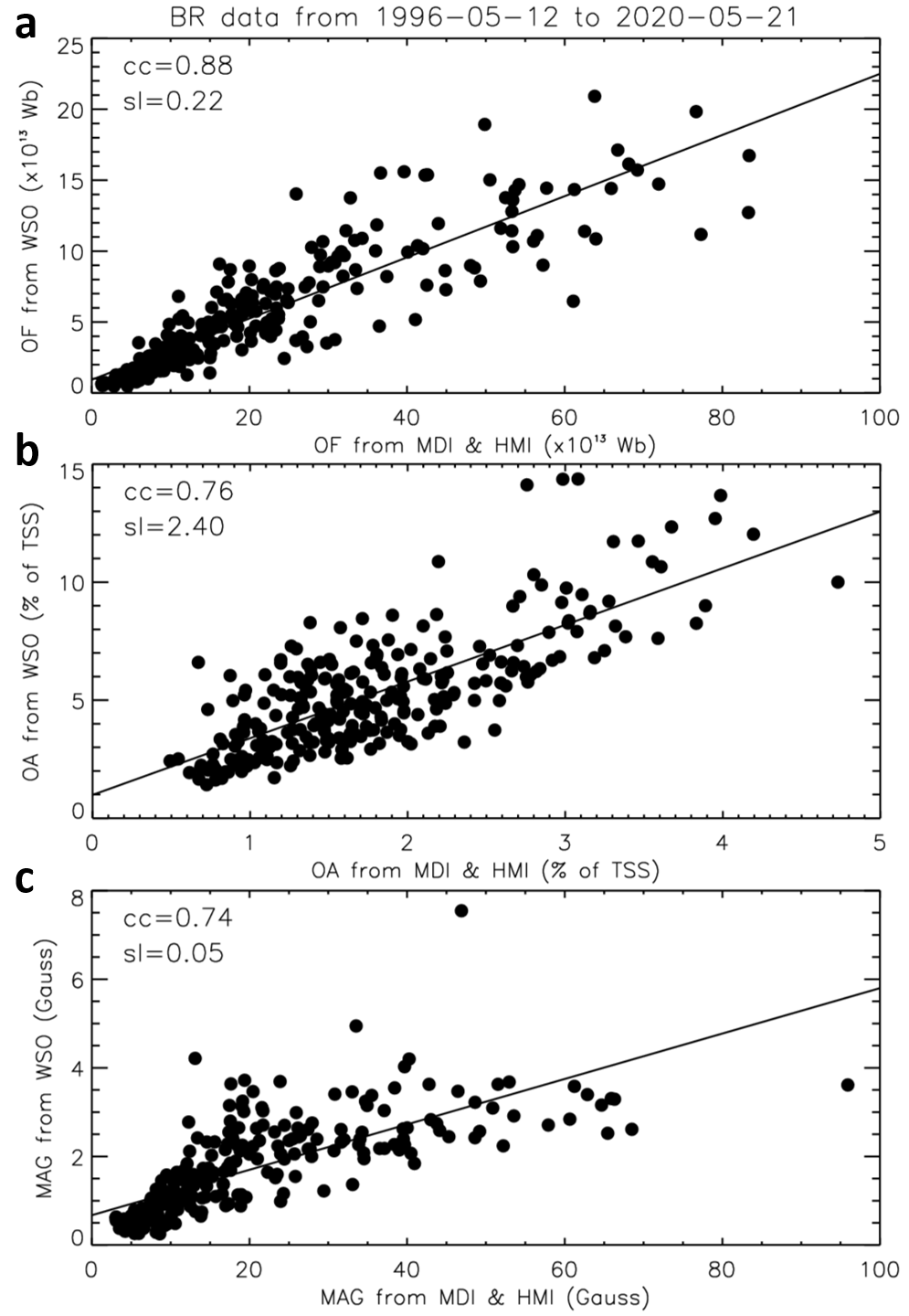}
	\caption{The scatter plots showing the correlation between the parameters derived based on the WSO data 
	and those based on the MDI and HMI data during 1996 -- 2020.
	}\label{fig:WSO_vs_MDI}
\end{center}
\end{figure*}

Here, we reduce the spatial resolution of the MDI and HMI data to 360 by 180, calculate the spherical harmonic coefficients up to 180 orders, and trace field lines at a 180 by 90 mesh to save the computing time.
These reduce synoptic charts are 30 time higher in resolution than WSO charts. Figure~\ref{fig:MDI_HMI} shows the 
solar cycle variations of the open area and open flux based on the MDI and HMI data. Their variation patterns look almost 
the same as those displayed in Figure~\ref{fig:open_field}, except that the absolute values are different. The open area
derived based on the WSO data is larger than that based on the MDI and HMI data, whereas the open flux 
based on the WSO data is smaller. The correlation between the WSO and MDI/HMI data is displayed in 
Figure~\ref{fig:WSO_vs_MDI}. It can be found that the slopes of the fitting lines obviously deviate 
from unity, suggesting that the open flux based on the WSO data is about 4-5 times smaller than
that based on the MDI and HMI data, the open area is about 2-3 times larger, and the averaged 
open field strength is about 20 times smaller. But the correlation coefficients of the open flux, open area and the open field strength are as high as
0.88, 0.76 and 0.74, respectively, suggesting the parameters derived from WSO data are well correlated with those from
the MDI and HMI data. Thus, the low spatial resolution of the WSO data should not distort the statistical results presented in the main text.

\bibliography{../../ahareference}{}
\bibliographystyle{aasjournal}



\end{document}